\documentclass[letterpaper, 10 pt, conference]{ieeeconf}  

\IEEEoverridecommandlockouts                              

\usepackage{graphicx}
\usepackage{tabu}
\usepackage{float}
\usepackage{amsmath} 
\usepackage{tikz}
\usetikzlibrary{matrix}
\usetikzlibrary{shapes.geometric, arrows}
\usepackage{algorithm}
\usepackage{algorithmic}
\usepackage{lipsum}
\usepackage{multicol}
\usepackage{balance}

\newcommand{\img}[1]{\begin{center}\includegraphics[width=\columnwidth]{{#1}}\end{center}}

\newcommand{\INDSTATE}[1][1]{\STATE\hspace{#1\algorithmicindent}}
\newcommand{\gridScale}{1.25}
\newcolumntype{M}[1]{>{\centering\arraybackslash}m{#1}}

\newcommand{\lesscaption}[1]{{\vspace{-0.75em}\caption{#1}\vspace{-1em}}}

\title{\LARGE \bf CDDT: Fast Approximate 2D Ray Casting for Accelerated Localization}

\author{Corey H. Walsh$^{1}$ and Sertac Karaman$^{2}$
\thanks{$^{1}$Corey H. Walsh is with the Department of Computer Science and Engineering, Massachusetts Institute of Technology,
        Cambridge, MA 02139, USA
        {\tt\small chwalsh@mit.edu}}%
\thanks{$^{2}$Sertac Karaman is with the Laboratory for Information and Decision Systems, Massachusetts Institute of Technology,
        Cambridge, MA 02139, USA
        {\tt\small sertac@mit.edu}}%
\thanks{This work was supported in part by the Office of Naval Research (ONR) through the ONR YIP program.}
}

\begin{document}

\maketitle
\thispagestyle{empty}
\pagestyle{empty}

\begin{abstract}

Localization is an essential component for autonomous robots. A well-established localization approach ~~~~~ combines ray casting with a particle filter, leading to a computationally expensive algorithm that is difficult to run on resource-constrained mobile robots. We present a novel data structure called the Compressed Directional Distance Transform for accelerating ray casting in two dimensional occupancy grid maps. Our approach allows online map updates, and near constant time ray casting performance for a fixed size map, in contrast with other methods which exhibit poor worst case performance. Our experimental results show that the proposed algorithm approximates the performance characteristics of reading from a three dimensional lookup table of ray cast solutions while requiring two orders of magnitude less memory and precomputation. This results in a particle filter algorithm which can maintain 2500 particles with 61 ray casts per particle at 40Hz, using a single CPU thread onboard a mobile robot.

\end{abstract}

\section{Introduction}

Determining a robot's location and orientation in a known environment, also known as localization, is an important and challenging problem in the field of robotics. Particle filters are a popular class of Monte Carlo algorithms used to track the pose of mobile robots by iteratively refining a set of pose hypotheses called particles. After determining an initial set of particles, the particle filter updates the position and orientation of each particle by applying a movement model based on available odometry data. Next, the belief in each particle is updated by comparing sensor readings to a map of the environment. Finally, the particles are resampled according to the belief distribution and the algorithm repeats.

While particle filters may be used for localization, they can be computationally expensive due to both the number of particles which must be maintained, and the evaluation of the sensor model. In robots with range sensors such as LiDAR or Sonar, ray casting is often used to compare sensor readings with the ground truth distance between the hypothesis pose and obstacles in a map (visualized in Fig. 1). Ray casting itself is a complex operation, often dependent on map occupancy or geometry, and an evaluation of the sensor model may require tens of ray casts. Many effective particle filters maintain thousands of particles, updating each particle tens of times per second. As a result, millions of ray cast operations may be resolved per second, posing a significant computational challenge for resource constrained systems. 

\begin{figure}[h!]
\vspace{6pt}
\img{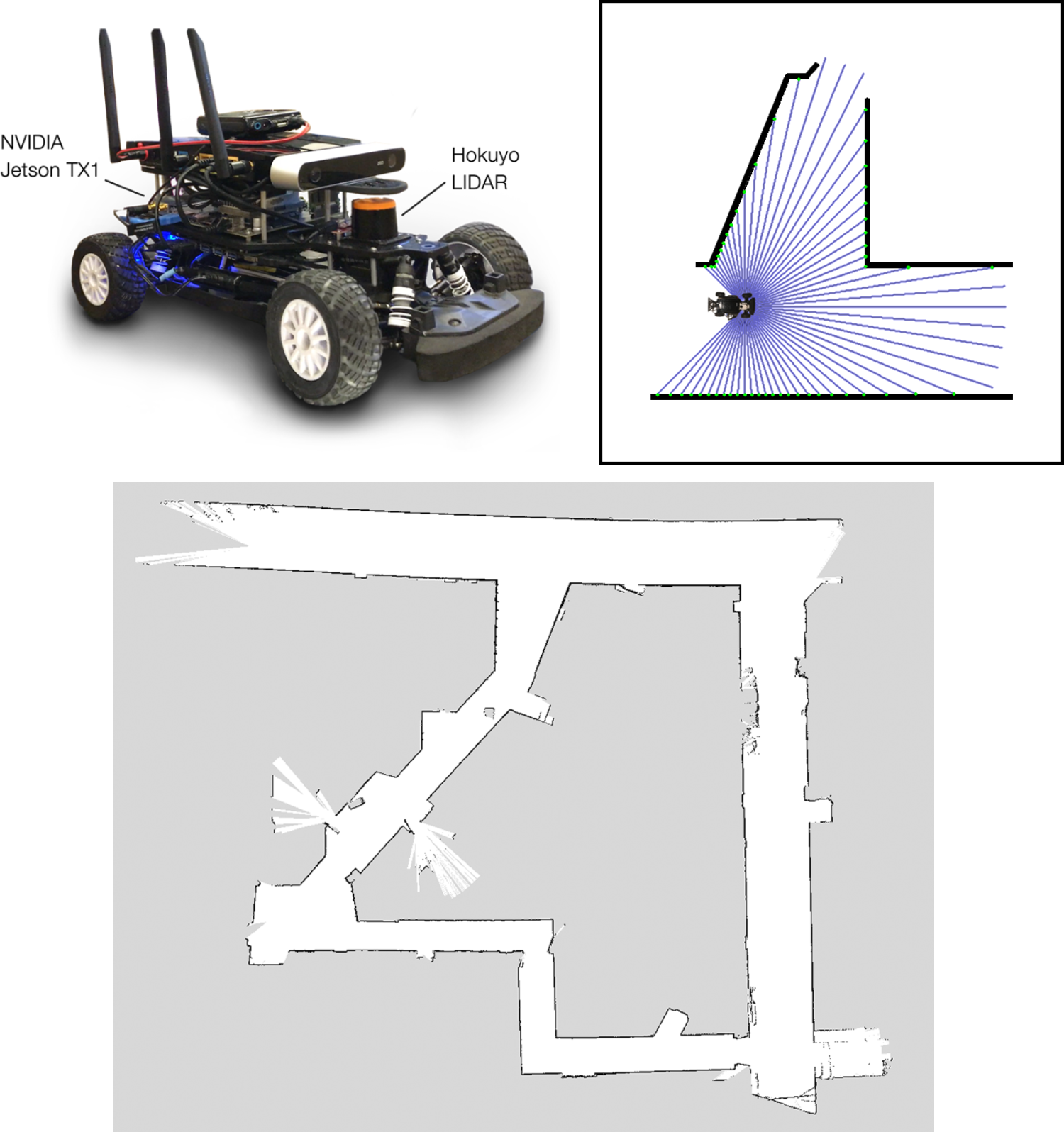}
\lesscaption{Robotic RACECAR mobile platform (top left). Synthetic occupancy grid map with the 61 ray cast queries used in the particle filter sensor model visualized in blue (top left). Stata basement occupancy grid map (bottom).}
\label{rays_car}
\end{figure}

Several well known algorithms exist for ray casting in two dimensional spaces such as Bresenham's Line (BL) algorithm \cite{bresenham} and ray marching (RM) \cite{raymarching}. Both algorithms work by iteratively checking points in the map starting at the query point and moving in the ray direction until an obstacle is discovered. This process does not provide constant performance because the number of memory reads depends on the distance to the nearest obstacle.

To combat the computational challenges of ray casting while localizing in a two-dimensional map, Fox et al. \cite{localization} suggest the use of a large three-dimensional lookup table (LUT) to store expected ranges for each discrete $(x,y,\theta)$ state. While this is simple to implement and does result in large speed improvements as compared to ray casting, it can be prohibitively memory intensive for large maps and/or resource constrained systems. In a 2000 by 2000 occupancy map, storing ranges for 200 discrete directions would require over 1.5GB. While this memory requirement may be acceptable in many cases, it scales with the area of the map - a 4000 by 4000 map would require over 6GB for the same angular discretization, which is larger than the random-access memory on-board many mobile robots.

The main contribution of this paper is a new acceleration data structure, called the Compressed Directional Distance Transform (CDDT) which allows near constant time two dimensional ray casting queries for an occupancy grid map of fixed size. The algorithm is benchmarked against several common ray casting methods. We provide an open-source implementation of CDDT and the other methods evaluated in a library called RangeLibc\footnote[3]{https://github.com/kctess5/range\_libc}. The CDDT algorithm has been applied to a particle filter localization algorithm\footnote[4]{https://github.com/mit-racecar/particle\_filter}, allowing 2500 particles to be maintained at 40Hz with 61 ray casts per particle on a NVIDIA Jetson TX1 embedded computer.

We observe two orders of magnitude improvement in memory consumption when compared to the lookup table method, with little sacrifice in computation time. Additionally, we observe a large speedup when compared to the other ray casting methods considered, with similar memory requirements. Unlike other accelerated methods considered, CDDT allows for incremental map modifications.

The paper is organized as follows. Section 2 discusses existing two dimensional ray casting methods. Section 3 introduces terminology used throughout the paper. Section 4 describes the new algorithm and its various optimizations. Section 5 gives a theoretical analysis of CDDT's asymptotic complexity. Section 6 describes our experimental results and comparisons. Finally, section 7 presents our conclusions.

\section{Related Work}

While we focus on occupancy grids, some researchers have explored non-grid map representations to manage computational complexity of particle filtering. One such representation is the vector map \cite{vector_map} which models permanent environmental features as a set of line segments. This representation can be useful because it is sparse, allows for both analytic ray casts and analytic observation model gradients. However, it can be more complex to work with than occupancy grids since it requires a methods to convert laser scans into a sparse set of line segments and recognize overlapping features. Additionally, the analytic ray casting method proposed in \cite{vector_map} scales with $O(n*log(n))$ where $n$ is the number of line segments maintained, which can be large if the environment is large or does not lend itself well to vector representation. 

Bresenham's line algorithm \cite{bresenham} is one of the most widely used methods for two dimensional ray casting in occupancy grids. The algorithm incrementally finds the set of pixels that approximate the trajectory of a ray starting from the query point $(x,y)_{query}$ and progressing in the $\theta_{query}$ direction one pixel at a time. The algorithm terminates once the nearest occupied pixel is found, and the Euclidean distance between that occupied pixel and $(x,y)_{query}$ is reported. This algorithm is widely used in particle filters due to its simplicity and ability to operate on a dynamic map. The primary disadvantage is that it is slow, potentially requiring hundreds of memory accesses for a single ray cast. While average performance is highly environment dependent, Bresenham's Line algorithm is linear in map size in the worst case.

\begin{figure}[h]
\vspace{-0.25em}
\img{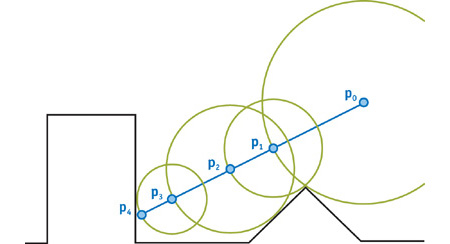}
\caption{Visualization of ray marching starting at $p_0$ towards $p_4$. Green circle around each query point represents the distance to the nearest obstacle from that point. Blue dots represent the query points, labeled in order of execution. From \cite{sphere:source}.}
\label{spheres}
\end{figure}

Similar to Bresenham's Line, the ray marching \cite{raymarching} algorithm checks points along the line radiating outwards from the query point until an obstacle is found. The primary difference is that ray marching makes larger steps along the query ray, thereby avoiding unnecessary memory reads. Beginning at $(x,y)_{query}$, the ray marching algorithm proceeds in the $\theta_{query}$ direction, stepping along the line by the minimum distance between each visited point and the nearest obstacle in any direction (graphically demonstrated in Fig. \ref{spheres}). The algorithm terminates when the query point coincides with an obstacle in the map. A precomputed Euclidean distance transform \cite{euclidean_distance_transforms} of the occupancy map provides the distance between visited points and the nearest obstacles.

Ray marching is on average faster than Bresenham's line, but edge cases exist in which the theoretically asymptotic runtime is equivalent. As noted by Zuiderveld et al.~\cite{acceleratedraymarching}, the traversal speed of rays rapidly decreases as sampled positions approach obstacles. For this reason, rays which travel parallel to walls progress slowly as compared to those passing through open areas. Thus, the performance of ray marching exhibits, roughly speaking, a long tail distribution as seen in (Fig. \ref{violin:basement:all}), i.e. a small number of queries take a disproportionately long time, which can be problematic for near real time algorithms.

As previously described, a common acceleration technique for two dimensional ray casting is to precompute the ray distances for every state in a discrete grid and store the results in a three dimensional LUT for later reference. Theoretically, the LUT approach has constant query runtime, though actual performance access pattern dependent as CPU caching effects are significant in practice.

State space discretization implies approximate results, since intermediate states must be rounded. While the effect of rounding query position is small, rounding $\theta_{query}$ may have significant effects since angular roundoff error accumulates with ray length. For queries $(x,y,\theta)_{query}$ discretized into $\lfloor(x,y,\theta)\rceil$, the distance between the end of the ray $\lfloor(x,y,\theta)\rceil$ and its projection onto the line implied by $(x,y,\theta)_{query}$ becomes large as the length of the ray increases. Rather than simply rounding to the nearest grid state, one may improve accuracy of queries which lie between discrete states by querying the neighboring states and interpolating the results. Interpolation requires extra computation since two ray casts must be performed, and is therefore slower than rounding to the nearest grid state. We do not perform interpolation, since as we demonstrate (Fig. \ref{error_vs_theta_discretization}), small errors have little impact on localization accuracy.






\section{Problem Formulation and Notation}

\begin{figure}[h!]
\begin{center}\includegraphics[width=7.6cm]{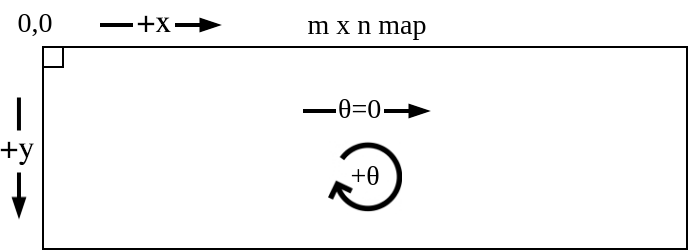}\end{center}
\vspace{-1em}
\caption{Occupancy grid map coordinate system}
\label{coordinate_system}
\end{figure}

We define the problem of ray casting in occupancy grids as follows. We assume a known occupancy grid map in which occupied cells have value 1, and unoccupied cells have value 0. Given a query pose $(x,y,\theta)_{query}$ in map space, the ray cast operation finds the nearest occupied pixel $(x,y)_{collide}$ which lies on the ray starting at the position $(x,y)_{query}$ pointing in the $\theta_{query}$ direction, and returns the Euclidean distance $d_{ray}$ between $(x,y)_{query}$ and $(x,y)_{collide}$.

$$ d_{ray} = \left\Vert\begin{pmatrix}
           x \\
           y 
         \end{pmatrix}_{query} - 
         \begin{pmatrix}
           x \\
           y
         \end{pmatrix}_{collide}\right\Vert_2
$$

We denote the discretized query pose as $\lfloor(x,y,\theta)_{query}\rceil$. A $\theta$ slice through the LUT is a 2D subset of the full 3D LUT in which the value of $\theta$ is held constant and $x,y$ are varied. The number of discrete $\lfloor\theta\rceil$ values is denoted $\theta_{discretization}$. Fig. \ref{coordinate_system} demonstrates our chosen coordinate system.

\section{The Compressed Directional Distance Transform Algorithm}

Although the three dimensional table used to store precomputed ray cast solutions in a discrete state space is inherently large, it is highly compressible. This is most apparent in the cardinal directions ($\theta=0,\frac{\pi}{2},\pi,\frac{3\pi}{2}$), in which adjacent values along a particular dimension of the table increase by exactly one unit of distance for unobstructed positions as in Fig. \ref{cardinal_example}. Our data structure is designed to compress this redundancy while still allowing for fast queries in near constant time. We accomplish this though through what we refer to as a Compressed Directional Distance Transform described here.

\begin{figure}[h]
\vspace{4pt}
\begin{center}
\begin{tikzpicture}
\fill[black,scale=\gridScale] (0.5,0.5) rectangle (1,1);
\fill[black,scale=\gridScale] (-0.5,0) rectangle (0,0.5);
\fill[black,scale=\gridScale] (0,-0.5) rectangle (0.5,0);
\fill[black,scale=\gridScale] (0.5,-1) rectangle (1,-0.5);
\draw[scale=\gridScale,step=0.5cm,color=white] (-1,-1.1) grid (1,1.1);
\draw[scale=\gridScale,step=0.5cm,color=gray] (-1,-1) grid (1,1);
\end{tikzpicture}
\begin{tikzpicture}
\draw[scale=\gridScale,step=0.5cm,color=white] (-1,-1.1) grid (1,1.1);
\draw[scale=\gridScale,step=0.5cm,color=gray] (-1,-1) grid (1,1);
\matrix[matrix of nodes,nodes={inner sep=0pt,text width=.5cm,align=center,minimum height=0.45cm, scale=\gridScale}]{
3 & 2 & 1 & 0 \\
1 & 0 & - & - \\
2 & 1 & 0 & - \\    
3 & 2 & 1 & 0\\};
\end{tikzpicture}
\end{center}
\vspace{-0.75em}
\caption{4x4 occupancy grid (left) and associated LUT slice for $\theta=0^{\circ}$ (right). Occupied grid cells are filled in black, while free space is white. The numbers in the LUT slice indicate distance in pixel units to the nearest occupied pixel in the rightward direction.}
\label{cardinal_example}
\end{figure}

As opposed to the Euclidean distance transform of an occupancy grid map, which stores the distance to the nearest obstacle in any direction $\theta$ for each possible grid state, what we call a directional distance transform (DDT) stores the distance to the nearest obstacle in a particular direction $\lfloor\theta\rceil$. In this sense, it is similar to a two dimensional slice of the LUT for a fixed angle.


The key difference between a single $\theta$ slice of the LUT and a DDT for the same $\theta$ is the way they are computed and indexed. To compute the LUT slice, rays are cast in the $\theta$ direction for every $\lfloor(x,y)\rceil$. At runtime, each $(x,y)_{query}$ is discretized to $\lfloor(x,y)_{query}\rceil$ and the LUT slice is read.

\begin{figure}[t!]
\img{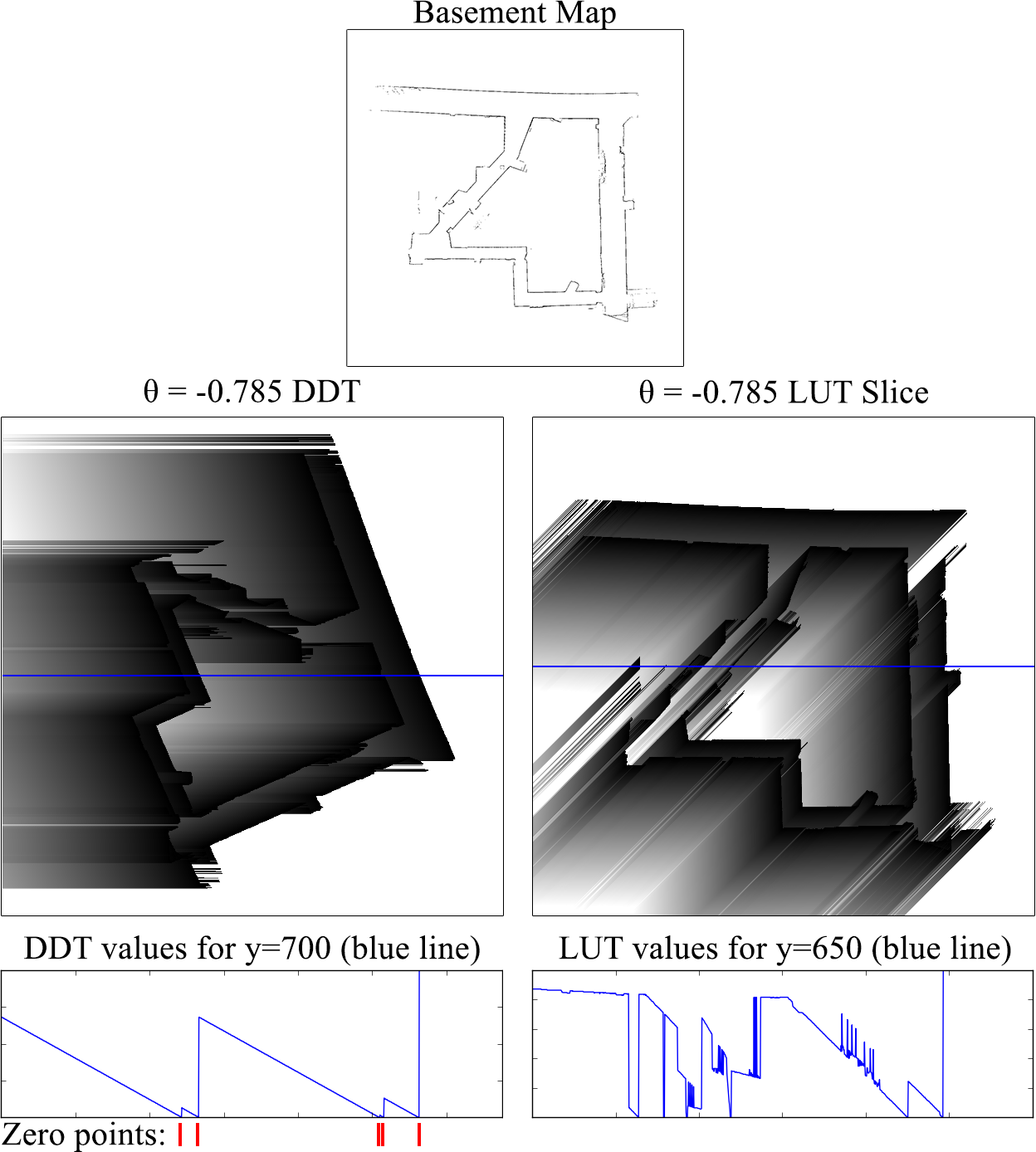}
\vspace{-0.75em}
\caption{Comparison of a DDT (middle left) and LUT (middle right) slice for the same value of $\theta$. Color encodes distance values, with black meaning 0 distance and white meaning large distance. Each row in the DDT is characterized by a sawtooth function (bottom left), whereas each row of the LUT slice is messy (bottom right). Notice that both images look similar, differing in rotation. In the DDT, scene geometry is rotated by a factor of $\theta$ about the origin and rays are cast along the $x$ axis. In contrast, the LUT slice is built by ray casting in the $\theta$ direction.}
\label{ddt_lut_comparison}
\vspace{-0.75em}
\end{figure}

In contrast, to compute the DDT, the obstacles in the map are rotated about the origin by $-\lfloor\theta\rceil$ and ray casting is implicitly performed in the $\theta=0$ direction, as demonstrated by Fig. \ref{ddt_lut_comparison}. During ray casting, rather than directly indexing the DDT using the query coordinates $(x,y,\theta)_{query}$ as is done with a LUT, one first applies the same rotation used to construct the DDT, and indexes the DDT using the transformed query coordinates $(x,y,\theta)_{rot}$. Thus, roughly the same operation is computed in both the DDT and the LUT, but while one changes the ray cast direction to populate the LUT, one rotates scene geometry and ray casts in a constant direction to populate the DDT.



The distinction between the LUT slice and the DDT may be subtle, but it has an important effect. Since ray casting is always performed in the $\theta=0$ direction to populate the DDT, all values in the same row of the DDT either increase by one unit with respect to their neighbor in the $\theta=0$ direction, or go to zero. Thus each row of the DDT may be characterized as a sawtooth function where the zero points correspond to obstacles in the map. This characterization as a sawtooth function provides a natural method of lossless compression: keep the zero points and discard the rest.

\begin{figure}[h]
\vspace{2pt}
\img{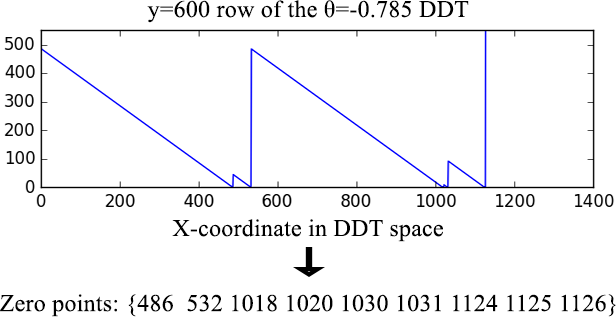}
\vspace{-0.75em}
\caption{Demonstration of compression from a sawtooth function to a list of zero points. Consecutively valued zero points exist when projected scene geometry spans greater than one unit along the x axis.}
\label{sawtooth_to_cddt}
\end{figure}


The conversion from DDT to CDDT slice is done by storing the $x$ coordinates of each zero point for every row of the DDT as demonstrated for a single row in Fig. \ref{sawtooth_to_cddt}. At query time, the sawtooth function (i.e. exactly the distance to the nearest obstacle in the query direction) encoded in the DDT may be quickly recovered by finding the nearest zero point in the correct row of the CDDT slice. We refer to the list of zero points for a single row as a CDDT bin. Similar to the LUT, the full CDDT is defined as every CDDT slice for all discrete values of $\theta$.

For performance, it is not necessary to compute or store the full DDT, rather, the CDDT may be directly constructed by projecting each obstacle into the coordinate space of the DDT for every $\lfloor\theta\rceil$ and storing its projected $x$ coordinate in the CDDT bin corresponding to its $y$ coordinate and $\lfloor\theta\rceil$ using a 3x3 projection matrix $P_{DDT_{\theta}}$. While the projection is primarily a rotation, in our implementation a translation is also applied to ensure that the projected $y$ coordinate is non-negative for use in indexing the correct bin. After all geometry has been projected into the CDDT, each bin is processed to facilitate later queries. The exact processing depends on the structure used for successor and predecessor queries. Not only does the direct construction of the CDDT greatly reduce the amount of memory required to store a lookup table, it also reduces precomputation time significantly since ray casting every discrete state is not necessary.

While conceptually simple, implementing the CDDT data structure construction and traversal routines requires careful consideration in order to capture all edge cases and to minimize unnecessary computation. In this sense, it is more complex to implement than the alternatives considered, however there are many opportunities for optimization which yield real-world speed up. To ease this burden, we provide our implementation$^3$ as well as Python wrappers under the Apache 2.0 license.

\subsection{Direct CDDT Construction Algorithm}
\vspace{-0.5em}
\begin{center}
\line(1,0){240}
\end{center}
\begin{algorithmic}
\STATE $edge\_map\gets map-morphological\_erosion(map)$
\STATE Initialize $\theta_{discretization}$ empty CDDT slices
\FOR {$\theta \in \{\lfloor\theta\rceil\}$} 
        \FOR{each occupied pixel $(x,y)\in edge\_map$}
            \STATE $(x,y)_{DDT_{\theta}} = P_{DDT_{\theta}}*\begin{pmatrix}
                        x \\
                        y \\
                        1
             \end{pmatrix}$
             \FOR {each CDDT bin overlapping with $y_{DDT_{\theta}}$}
                \STATE bin.append($x_{DDT_{\theta}}$)
             \ENDFOR
        \ENDFOR
        \FOR{each CDDT bin}
            \STATE {\color{gray}\# Initialize successor/predecessor structure}
            \STATE bin = initialize\_bin\_structure(bin)
        \ENDFOR
\ENDFOR
\end{algorithmic}
\begin{center}
\vspace{-0.75em}
\line(1,0){240}
\end{center}

\subsection{CDDT Query Algorithm}
\vspace{-0.5em}
\begin{center}
\line(1,0){240}
\end{center}
\begin{algorithmic}
\STATE \textbf{function ray\_cast($(x,y,\theta)_{q})$}
\INDSTATE[1] $(x,y)_{DDT_{\theta_{q}}} = P_{DDT_{\theta_{q}}}*\begin{pmatrix}
                        x_{q} \\
                        y_{q} \\
                        1
             \end{pmatrix}$
\INDSTATE[1] bin $\gets$ zero points in row $y_{DDT_{\theta_{q}}}$ of CDDT slice $\theta_{q}$
\INDSTATE[1] $x_{collide}$ = smallest element $x_{collide} > x_{DDT_{\theta_q}}\in$ bin
\INDSTATE[1] \textbf{return} abs($x_{DDT_{\theta_q}} - x_{collide}$)
\STATE \textbf{end function}\end{algorithmic}
\begin{center}
\vspace{-0.75em}
\line(1,0){240}
\end{center}

The discovery of $x_{collide}$ in a given CDDT bin is a successor or predecessor query on the bin data structure. We have experimented with both sorted vectors and B-trees. While both structures offer logarithmic successor and predecessor queries with similar performance in practice, the B-tree also offers logarithmic time insertions and deletions, which is useful for incremental modification.


\subsection{Further Optimizations} \label{optimizations}

\begin{figure}[h]
\vspace{5pt}

\img{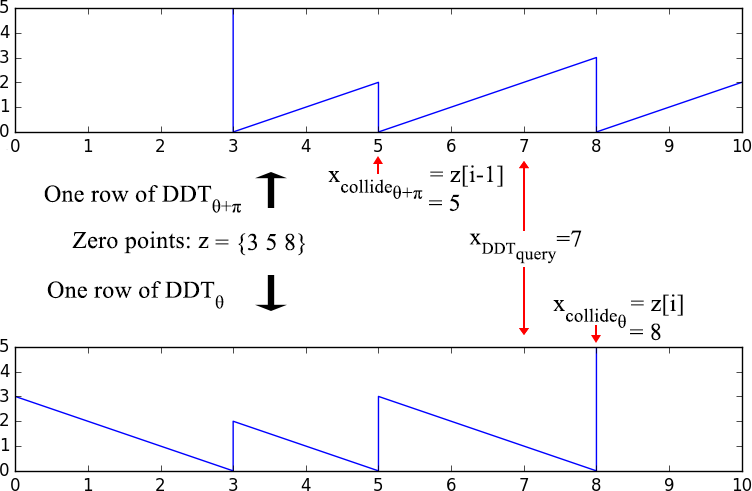}
\vspace{-0.75em}

\caption{Demonstration of reconstructing rows of two DDT slices from a single set of zero points.}
\vspace{-1em}

\label{bidirectional_zero_points}
\end{figure}

Extracting zero points from each row of the DDT introduces rotational symmetry. From one set of zero points, two rows of the DDT can be reconstructed - the rows corresponding to $\theta$ and $\theta+\pi$ for any particular $\theta$ (as shown in Fig. \ref{bidirectional_zero_points}). Therefore, one only need compute and store CDDT slices for $0\leq\lfloor\theta\rceil<\pi$ and the DDT for $0\leq\lfloor\theta\rceil<2\pi$ may be inferred by reversing the binary search direction, resulting in a factor of two reduction in memory usage.

Rotational symmetry may be further exploited in scenarios where ray casts are performed radially around a single point. While traversing the data structure to resolve a ray cast query $(x,y,\theta)$, the ray cast for $(x,y,\theta+\pi)$ may be resolved with a single additional memory read. Once a search algorithm is used to discover the index $i$ of $x_{collide_{\theta}}$ in the CDDT, the index of $x_{collide_{\theta+\pi}}$ is simply $i-1$ as in Fig. \ref{bidirectional_zero_points}. For example, this symmetry can be used in robots with laser scanners sweeping angles larger than $180^{\circ}$ to reduce the number of data structure traversals required to compute the sensor model by up to a factor of two. 

With small modifications, another factor of two reduction in CDDT size could be attained by storing zero points as 16 bit integers rather than 32 bit floats with an acceptable precision loss for particle filtering applications.

\begin{figure}[h]
\begin{center}\includegraphics[width=6.5cm]{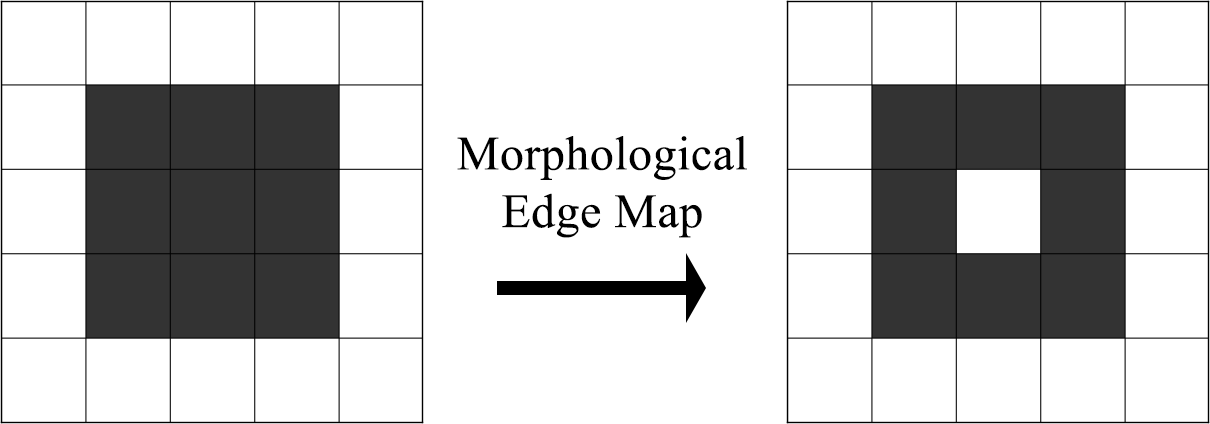}\end{center}
\vspace{-0.75em}
\caption{Example map left, edge map right.}
\label{edge_map}
\vspace{-0.75em}
\end{figure}

By removing entries in the CDDT which can never possibly result in a ray collision, it is possible to further reduce the memory footprint of the data structure. Consider a 3x3 block of obstacles in an otherwise empty map as in Fig. \ref{edge_map}. The center obstacle will never be the nearest neighbor in any ray casting query, because any such query would first intersect with one of the surrounding obstacles. To exploit this, one can use the edge map for CDDT generation without loss of generality. To ensure correct results with this optimization, one must check if the query point overlaps with an obstacle to avoid ray casting from the middle of removed obstacles.


Additionally, consider a line of obstacles aligned along the X-axis. Every element in this line will be projected into a single zero point bin in the $\theta=0$ CDDT slice. However, the middle elements of the line will never result in a collision. Any ray cast from points on the line of obstacles will return early in the occupancy grid check, and any ray cast from non-overlapping points co-linear in the $\theta$ or $\theta+\pi$ directions will intersect either the first or last obstacle in the line. Therefore in the $\theta=0$ CDDT slice, one may discard the zero points corresponding to the middle elements without introducing error. This form of optimization is simple to compute in the cardinal directions, but non-trivial for arbitrary $\lfloor\theta\rceil$ not aligned with an axis. Rather than analytically determining which obstacles may be discarded, it is simpler to prune the data structure by ray casting from every possible state, discarding any unused zero point. 

Pruning does increase pre-computation time. However, the reduction of memory usage is worthwhile for static maps (see Fig. \ref{table:synthetic:init}). In addition to memory reduction, we find that pruning slightly improves runtime performance, likely as a result of improved caching characteristics and the reduced number of zero points. We refer to the pruned datastructure as PCDDT.

\subsection{Incremental CDDT Obstacle Modification Algorithm}
\vspace{-0.5em}
\begin{center}
\line(1,0){240}
\end{center}
\begin{algorithmic}
\STATE \textbf{function update\_obstacle(int $x,y,$ bool $occupied$)}
\INDSTATE[1] map[$x$][$y$] $\gets occupied$ 
\INDSTATE[1] \textbf{for} $\theta \in \{\lfloor\theta\rceil\}$  \textbf{do}
\INDSTATE[2] $(x,y)_{DDT_{\theta}} = P_{DDT_{\theta}}*\begin{pmatrix}
                        x \\
                        y \\
                        1
             \end{pmatrix}$
\INDSTATE[2] \textbf{for} each CDDT bin overlapping with $y_{DDT_{\theta}}$ \textbf{do}
\INDSTATE[3] \textbf{if} $occupied$ \textbf{do}
\INDSTATE[4] bin.insert($x_{DDT_{\theta}}$)
\INDSTATE[3] \textbf{else do}
\INDSTATE[4] bin.remove($x_{DDT_{\theta}}$)
\INDSTATE[3] \textbf{endif}
\INDSTATE[2] \textbf{end for}
\INDSTATE[1] \textbf{end for}
\STATE \textbf{end function}\end{algorithmic}
\begin{center}
\vspace{-0.75em}
\line(1,0){240}
\end{center}

Since each element in the scene corresponds to a predictable set of zero points in the CDDT, if the map changes, one may insert or remove zero points accordingly, as outlined by algorithm D. For efficiency, we recommend using a B-tree data structure to store zero points. B-trees provide asymptotically logarithmic insertion, deletion, and query runtime, as well as good cache characteristics in practice.

If one uses the morphological pre-processing steps during initial CDDT construction, special care must be taken in the incremental obstacle deletion routine. Specifically, if edge obstacles are removed, previously occluded non-edge pixels may be revealed, and must therefore be inserted in order to retain data structure consistency. This process is not prohibitively expensive since it only requires checking the eight adjacent pixels. Incremental map modifications are generally incompatible with PCDDT, as the exhaustive pruning operation makes ensuring data structure consistency during obstacle deletion non-trivial.

\section{Analysis}

In this section, we refer to the width and height of the source occupancy grid map as $w$ and $h$, respectively. We refer to the diagonal length across the map as $d_{w,h} = \sqrt{w^2+h^2}$. The CDDT algorithm requires the original map data to check for overlaps between $(x,y)_{query}$ and obstacles prior to searching CDDT bins. Additionally, for each occupied map pixel, a total of $\theta_{discretization}$ float values are stored in the CDDT bins. Thus, the memory usage of the CDDT data structure is $O(n*\theta_{discretization}+w*h)$ where $n$ is the number of occupied pixels in the edge map. Since we must sort each bin after CDDT construction, pre-computation time is at worst $O(n*\theta_{discretization} + d_{w,h}^2*\theta_{discretization}*log(d_{w,h}))$ for the same definition of $n$. In practice it is closer to $n*\theta_{discretization} + d_{w,h}*\theta_{discretization}$ since each CDDT bin has a small number of elements on average, as evidenced by the high demonstrated compression ratio.

The pruning operation described in \ref{optimizations} reduces memory requirement, with a computational cost of $O(w*h*\theta_{discretization}*O(calc\_range)_{CDDT})$. The precise impact of pruning on memory usage is scene dependent, and difficult to analyze in the general case.

The ray cast procedure has three general steps: projection into CDDT coordinate space, the search for nearby zero points, and the computation of distance given the nearest zero point. The first and last steps are simple arithmetic, and therefore are theoretically constant time. The second step requires a successor or predecessor query on the CDDT bin structure. As previously discussed, the number of zero points in each CDDT bin tends to be small and is bounded in map size. Thus, at worst the search operation using either a sorted vector or B-tree requires $\log{(d_{w,h})}$ which is a small constant value for a fixed size map. Therefore, for a given map size, our algorithm provides $O(1)$ query performance.

When using a B-tree for zero points, the cost of toggling an occupancy grid cell's state is $O(\theta_{discretization}\log{k})$ where $k$ is the number of elements in each associated CDDT bin. Using the same argument of bin size boundedness for fixed size maps, the cost of this update becomes $O(\theta_{discretization})$ for maps of fixed dimension. In any case, this cost is generally not prohibitive for real-time performance in dynamic maps for reasonable choice of $\theta_{discretization}$ (Fig. \ref{error_vs_theta_discretization}).

\section{Experiments}

We have implemented the proposed algorithm in the C++ programming language, as well as Bresenham's Line, ray marching, and the LUT approach for comparison. Our source code$^1$ is available for use and analysis, and Python wrappers are also provided. All synthetic benchmarks were performed on a computer with an Intel Core i5-4590 CPU @ 3.30GHz with 16GB of 1333MHz DDR3 ram, running Ubuntu 14.04.

\begin{figure}[h!]
\vspace{6pt}
\begin{center}
\begin{tabular}{ | m{8em} | m{2.3cm}| m{1.5cm} | } 
\hline
 \multicolumn{3}{|c|}{Basement Map, $\theta_{discretization}$: 108} \\
 \hline
 Method & Memory Usage & Init. Time \\
 \hline
 Bresenham's Line & 1.37 MB & 0.006 sec  \\
 Ray Marching & 5.49 MB & 0.16 sec  \\
 CDDT & 6.34 MB & 0.067 sec  \\
 PCDDT & 4.07 MB & 2.2 sec  \\
 Lookup Table & 296.63 MB & 15.3 sec  \\
\hline \hline
 \multicolumn{3}{|c|}{Synthetic Map, $\theta_{discretization}$: 108} \\
 \hline
 Method & Memory Usage & Init. Time \\
 \hline
 Bresenham's Line & 1 MB & 0.004 sec  \\
 Ray Marching  & 4 MB & 0.13 sec  \\
 CDDT & 2.71 MB & 0.03 sec  \\
 PCDDT & 1.66 MB & 0.74 sec  \\
 Lookup Table & 216 MB & 9.1 sec  \\
\hline
\end{tabular}
\end{center}
\caption{Construction time and memory footprint of each method. Ranges stored in lookup table with 16 bit integers, initialized with ray marching.}
\label{table:synthetic:init}
\end{figure}

\begin{figure}[h!]
\img{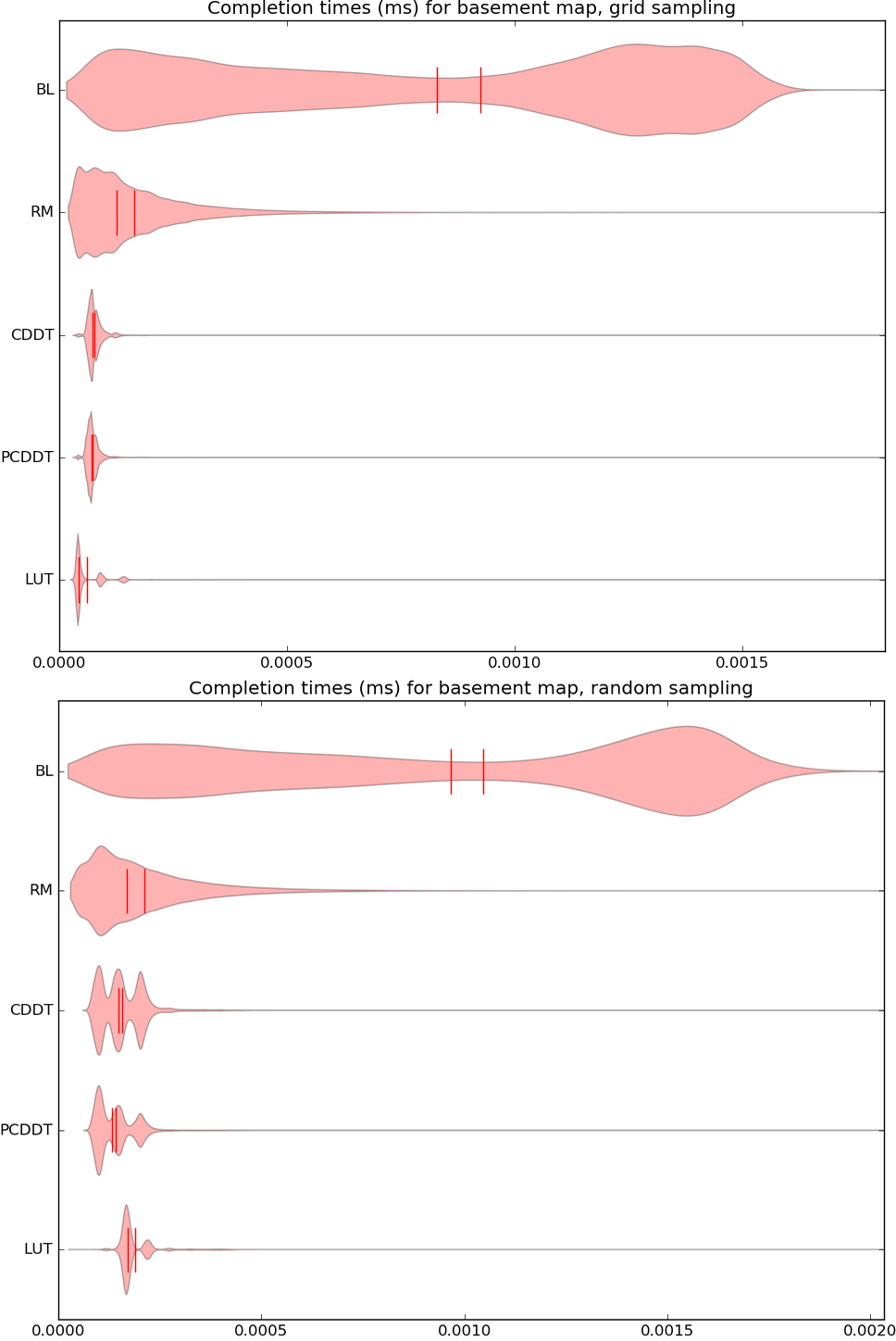}
\caption{Violin plots demonstrating histogram of completion time over a large number of queries for each ray cast method. Basement map. X axis shows time in milliseconds, and Y axis shows the number of queries that completed after that amount of time.}
\label{violin:basement:all}
\vspace{0.5em}
\end{figure}

We evaluate algorithm performance in two synthetic benchmarks, using two maps. The first "grid" benchmark computes a ray cast for each point in a uniformly spaced grid over the three dimensional state space. The second "random" benchmark performs a many ray casts for states generated uniformly at random. The so called Synthetic map (Fig. \ref{rays_car}) was created with Adobe Photoshop, whereas the basement map (Fig. \ref{rays_car}) was created via a SLAM algorithm on the RACECAR platform \footnote[5]{http://racecar.mit.edu} while navigating the Stata basement. 

\begin{figure}[h]
\begin{center}
\vspace{6pt}
\newcolumntype{C}[1]{>{\centering\arraybackslash}m{#1}}
\begin{tabular}{ | C{1.15cm} | C{1.25cm}| C{1.25cm} | C{1.24cm} | C{1.23cm}|  } 
\hline
\multicolumn{5}{|c|}{\textbf{Synthetic Map Ray Cast Benchmarks}} \\
\hline \hline
\multicolumn{5}{|c|}{Random Sampling} \\
\hline
Method & Mean & Median & IQR & Speedup \\
\hline
BL & 1.19e-06 & 1.41e-06 & 7.71e-07 & 1 \\
RM & 1.52e-07 & 1.25e-07 & 1.05e-07 & 7.81 \\
CDDT & 1.24e-07 & 1.05e-07 & 5.3e-08 & 9.59 \\
PCDDT & 1.19e-07 & 1.01e-07 & 5e-08 & 10.02 \\ 
LUT & 1.82e-07 & 1.68e-07 & 1.4e-08 & 6.55 \\
\hline \hline
\multicolumn{5}{|c|}{Grid Sampling} \\
\hline
Method & Mean & Median & IQR & Speedup \\
\hline
BL & 1.03e-06 & 1.20e-06 & 6.79e-07 & 1 \\ 
RM & 1.27e-07 & 1.03e-07 & 1.06e-07 & 8.06 \\
CDDT & 7.02e-08 & 6.8e-08 & 1e-08 & 14.63 \\
PCDDT & 6.94e-08 & 6.8e-08 & 9e-09 & 14.80 \\
LUT & 6.33e-08 & 4.2e-08 & 4.6e-08 & 16.21 \\
\hline
\end{tabular}
\label{table:synthetic:benchmarks}
\end{center}
\end{figure}
\begin{figure}[h]
\begin{center}
\newcolumntype{C}[1]{>{\centering\arraybackslash}m{#1}}
\begin{tabular}{ | C{1.15cm} | C{1.25cm}| C{1.25cm} | C{1.24cm} | C{1.23cm}|  } 
\hline
\multicolumn{5}{|c|}{\textbf{Basement Map Ray Cast Benchmarks}} \\
\hline \hline
Method & Mean & Median & IQR & Speedup \\
\hline
\multicolumn{5}{|c|}{Random Sampling} \\
\hline
BL & 9.66e-07 & 1.05e-06 & 1.08e-06 & 1 \\
RM & 2.12e-07 & 1.68e-07 & 1.64e-07 & 4.56 \\
CDDT & 1.58e-07 & 1.49e-07 & 9.1e-08 & 6.13 \\
PCDDT & 1.41e-07 & 1.32e-07 & 6.5e-08 & 6.83 \\
LUT & 1.89e-07 & 1.7e-07 & 2.1e-08 & 5.10 \\
\hline \hline
\multicolumn{5}{|c|}{Grid Sampling} \\
\hline
Method & Mean & Median & IQR & Speedup \\
\hline
BL & 8.29e-07 & 9.24e-07 & 9.53e-07 & 1 \\
RM & 1.65e-07 & 1.26e-07 & 1.34e-07 & 5.02 \\
CDDT & 7.69e-08 & 7.2e-08 & 1.6e-08 & 10.78 \\
PCDDT & 7.32e-08 & 7e-08 & 1.4e-08 & 11.33 \\
LUT & 6.13e-08 & 4.3e-08 & 4.6e-08 & 13.53 \\
\hline
\end{tabular}
\caption{Synthetic benchmark ray cast query runtime statistics for the Synthetic map (top) and Basement map (bottom). All times listed in seconds, speedup relative to Bresenham's Line}
\label{table:basement:benchmarks}
\end{center}
\vspace{-1em}
\end{figure}

\begin{figure}[t!]
\vspace{6pt}
  \img{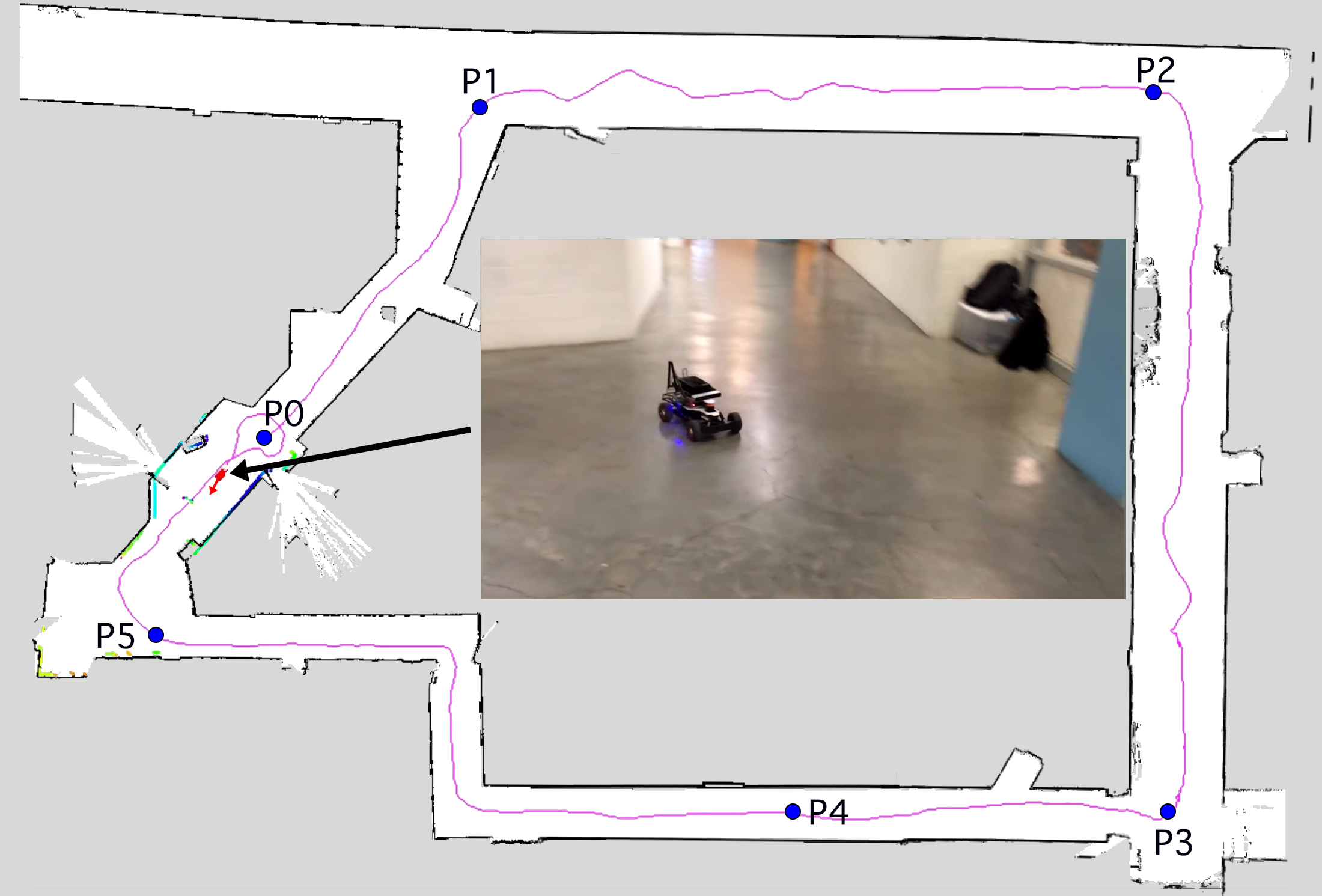}
  \vspace{-1em}
  \caption{Localization trail for a two minute sample of manually driving the RACECAR around the stata basement loop. Positions marked in blue correspond to markers on Fig. \ref{basement_mcl_runtime}. Small red arrow shows car's end position, corresponding to the image overlaid in the map. The map shown is a version of the Basement map modified for real-world accuracy.}
  \label{basement_mcl_slime}
\end{figure}

\begin{figure}[h]
\vspace{5pt}
\img{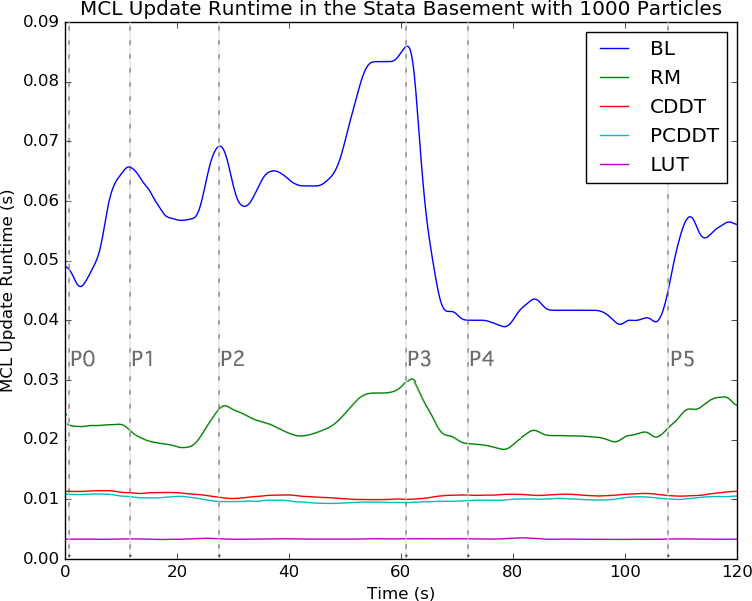}

\caption{Time required by the MCL update step of the particle filter algorithm with different ray casting methods over a two minute trial on the RACECAR. In all cases, the particle filter maintains 1000 particles, resolving 61 ray casts per particle. $\theta_{discretization}=120$ where applicable. Real-world position of the car at times marked P0-P5 shown in Fig. \ref{basement_mcl_slime}.}

\label{basement_mcl_runtime}
\end{figure}
\begin{figure}[h!]
\begin{tabular}{ | M{3.5em} | M{3.5em} | M{3.5em} | M{3.85em}| M{3.5em} | } 
 \hline
 \multicolumn{5}{|c|}{Max particles maintained at 40Hz with 61 rays/particle} \\
 \hline
  BL & RM & CDDT & PCDDT & LUT \\
 \hline
 400 & 1000 & 2400 & 2500 & 8500 \\
\hline
\end{tabular}
\caption{Maximum number of particles that can be maintained in real time (approx. 40Hz) on the NVIDIA Jetson TX1. Stata basement map, 61 ray casts per particle, $\theta_{discretization} = 120$ where applicable.}
\vspace{-1.5em}
\label{table:particle_filter:particles}
\end{figure}

To demonstrate real world performance, we have implemented$^4$  the particle filter localization algorithm using a beam mode sensor model. We provide information about the ray cast performance of each algorithm while being used to compute the sensor model (Fig. \ref{basement_mcl_runtime}), and the maximum number of particles each method was able to support in real time (Fig. \ref{table:particle_filter:particles}). In all particle filter benchmarks, we use the NVIDIA Jetson TX1 embedded computer onboard the RACECAR platform. We use a single thread for computing the Monte Carlo update step, though it could be easily parallelized across multiple threads for additional performance. 

Our sensor model is designed for the Hokuyo UST-10LX lidar scanner used aboard the RACECAR platform, which features a $270^{\circ}$ field of view. Since this FOV is in excess of $180^{\circ}$, we exploit radial symmetry discussed in subsection \ref{optimizations} to simultaneously ray cast in the $\theta$ and $\theta+\pi$ direction when possible. This optimization reduces the number of data structure traversals required by a third, while still resolving the same number of ray casts. As is standard practice, we down-sample the laser scanner resolution to reduce the number of ray casts per sensor model evaluation, and to make the probability distribution over the state space less peaked. We evaluate the sensor model for every scan received.

During particle filter execution, we track the amount of time the Monte Carlo update step takes, including the particle resampling, motion model, and sensor model steps. Fig. \ref{basement_mcl_runtime} demonstrates the MCL execution time over a two minute dataset collected on the RACECAR while driving the Stata basement loop. We find that the BL and RM cause in significant variance in sensor model execution time, depending on the location of the car, and the nearby map geometry. Specifically, BL and RM tend to be fast in constrained environments such as narrow hallways where beams are short on average, and slow in wide open area where beams are long on average (Fig. \ref{basement_mcl_slime},  \ref{basement_mcl_runtime}). In contrast, with the CDDT or LUT based methods, the MCL update step has very little location dependent variance in execution time. This finding is in line with our expectations given the long tail performance of BL and RM, and the theoretically near constant time nature of CDDT and LUT. In a real time setting, algorithms with highly variable runtimes are problematic, as one must budget for the worst-case execution time.

It is interesting to note that LUT provides very fast performance in the particle filter, roughly 3.4 times faster than PCDDT. We believe this is due to the tightly clustered memory access pattern of a well-localized particle filter, which yields a good low-level cache hit rate.

\begin{figure}[h]
\vspace{-4pt}
\img{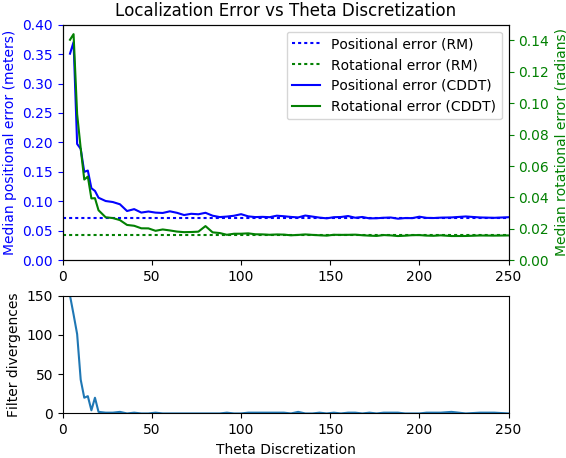}
\vspace{-0.85em}
\caption{The effect of the $\theta_{discretization}$ parameter on median positional and rotational localization error (top), and the number of particle filter divergences (bottom) during a five minute period of driving in a motion capture system. All other parameters are held constant. Notice that a $\theta_{discretization}$ above around 100 bins results in error characteristics similar to the ray marching algorithm.}
\label{error_vs_theta_discretization}
\end{figure}

To evaluate the effect of theta discretization on localization performance, we have used a motion capture system to gather ground truth state information for comparison with the state inferred by our particle filter. We autonomously drove the RACECAR around our motion capture environment for a period of five minutes while collecting all sensor data. Offline, we repeatedly performed particle filter localization on all collected data using a varied theta discretization parameter while tracking state inference error. Our results (Fig. \ref{error_vs_theta_discretization}) indicate that the approximate nature of ray casting methods using a discrete theta space (including LUT and CDDT) does not have a large impact on localization quality above a certain threshold.

Our testing framework provided a ground truth pose to the particle filter when the particle distribution significantly diverged from the motion capture data. This solution to the kidnapped robot problem \cite{localization} allowed us to test using extremely coarse $\theta$ discretizations. We present the number of times the ground truth position was provided in Fig. \ref{error_vs_theta_discretization}. In practice, some form of global localization could be used to recover from divergences, such as in Mixture-MCL \cite{localization_mixture_mcl}.

\section{Conclusions}

\balance

This work demonstrates that the proposed CDDT algorithm may be used in mobile robotics to accelerate sensor model computation when localizing in a two dimensional occupancy grid map. Unlike all methods considered other than Bresenham's Line algorithm, our method allows obstacles to be efficiently added or removed from the data structure without requiring full recomputation.

While the precomputed LUT method appears 1.1 to 3.4 times faster than CDDT, the memory and precomputation time required by the LUT is approximately two orders of magnitude larger (Fig. \ref{table:synthetic:init}, \ref{table:basement:benchmarks}). Compared to ray marching, CDDT is generally 1.2 to 2.4 times faster, with a more consistent query runtime. The comparison with the widely used Bresenham's Line algorithm is more stark, with CDDT providing a speedup factor of 6.8 to 14.8 in our benchmarks.

We showed that the use of a discrete theta space does not have a significant adverse effect on localization error, assuming a sufficient discretization is chosen. This result hold for both the proposed algorithm, and more traditional approaches such as the precomputed LUT.

While our experiments reveal a multi-modal distribution of completion times for CDDT ray cast queries, the vast majority of queries complete within a small constant factor of the median completion time, consistent with our asymptotic runtime analysis. We suspect the various modes in completion time are due to short circuit cases in our implementation where the query algorithm can return early, as well as caching effects.



\addtolength{\textheight}{20cm}
\clearpage

\end{document}